\documentclass[amsmath,amssymb,groupedaddress,superscriptaddress,lengthcheck,aps,prl] {revtex4}

\usepackage{graphics}
\usepackage{graphicx}
\usepackage{bm}
\usepackage[latin1]{inputenc}
\usepackage{textcomp}

\begin{document}

\title{Observation of non-conventional spin waves in composite fermion ferromagnets}


\author{U. Wurstbauer}
    \email{uw2106@columbia.edu }
    \affiliation{Department of Physics, Columbia University, New York, New York 10027, USA}

\author{D. Majumder}
    \affiliation{Department of Theoretical Physics, Indian Association for the Cultivation of Science,
    Jadavpur, Kolkata 700 032, India}

\author{S. S. Mandal}
    \affiliation{Department of Theoretical Physics, Indian Association for the Cultivation of Science,
    Jadavpur, Kolkata 700 032, India}

\author{I. Dujovne}
    \affiliation{Chemistry Department, University of Massachusetts, Amherst Massachusetts 01003}

\author{T. D. Rhone}
    \affiliation{Department of Physics, Columbia University, New York, New York 10027, USA}

\author{B. S. Dennis}
    \affiliation{Department of Physics and Astronomy, Rutgers University, Piscataway, New Jersey 08854, USA}

\author{A. F. Rigosi}
    \affiliation{Department of Physics, Columbia University, New York, New York 10027, USA}

\author{J. K. Jain}
    \affiliation{104 Davey Laboratory, Physics Department, Pennsylvania State University,
    University Park, Pennsylvania 16802, USA}

\author{A. Pinczuk}
    \affiliation{Department of Physics, Columbia University, New York, New York 10027, USA}
    \affiliation{Department of Applied Physics and Applied Math, Columbia University, New York, New York 10027, USA}

\author{K. W. West}
    \affiliation{Physics Department, Princeton University, Princeton, New Jersey 08544, USA}

\author{L. N. Pfeiffer}
    \affiliation{Physics Department, Princeton University, Princeton, New Jersey 08544, USA}

\date{\today}
\begin{abstract}
We find unexpected low energy excitations of fully spin-polarized composite-fermion ferromagnets in the fractional quantum Hall liquid, resulting from a complex interplay between a topological order manifesting through new energy levels and a magnetic order due to spin polarization. The lowest energy modes, which involve spin reversal, are remarkable in displaying unconventional negative dispersion at small momenta followed by a deep roton minimum at larger momenta. This behavior results from a nontrivial mixing of spin-wave and spin-flip modes creating a spin-flip excitonic state of composite-fermion particle-hole pairs.  The striking properties of spin-flip excitons imply highly tunable mode couplings that enable fine control of topological states of itinerant two-dimensional ferromagnets.
\end{abstract}


\maketitle 

Strong repulsive interaction between charged particles confined in two dimensions and subjected to a perpendicular magnetic field leads to the striking correlation phenomenon of the fractional quantum Hall effect (FQHE) \cite{Tsui82}. FQHE is explained by the emergence of composite fermions (CFs), bound states of electrons and an even number $2n$ of quantized vortices \cite{Jain89}. Composite fermions have a topological character because of the attached vortices, each of which produces a Berry phase of 2$\pi$ for a closed loop around it. The Berry phases partly cancel the effect of the perpendicular external field $B_{\perp}$ to produce an effective magnetic field $B_{\rm eff} = B_{\perp} - 2n\rho\phi_{0}$, where $\rho$ is the particle density and $\phi_{0} = hc/e$. As a result, composite fermions form Landau-like levels, called "$\Lambda$ levels" ($\Lambda$L's), in the reduced effective magnetic field. The splitting of the lowest Landau-level (LL) of electrons into $\Lambda$L's is thus a direct manifestation of the topological order of the FQHE. The main sequences of FQHE states at LL fillings $\nu = p/(2np \pm 1)$ (the integer $p$ is the number of filled $\Lambda$L's) are successfully described as integer quantum Hall effect of noninteracting CFs. We show below that the existence of $\Lambda$L's has nontrivial consequences for the physics of magnetism arising from the CF spin degree of freedom.\par
The FQHE states with $\nu < 1$ of the lowest LL are fully spin-polarized ferromagnets for a broad range of Zeeman energy and electron density. The basic physics of 2D ferromagnets entails a low energy spin wave (SW), for which only the orientation of spin changes \cite{Kallin}. In the small wave vector limit ($q \rightarrow 0$), Larmor's theorem stipulates that the SW energy is precisely equal to the bare Zeeman energy $E_{\rm Z} = g\mu_{B}B_{\rm tot}$, where $g$ is the Land\'e factor, $\mu_{B}$ is the Bohr magneton, and $B_{\rm tot}$ the total applied magnetic field. For a conventional ferromagnet, such as the one at $\nu =1$ \cite{Kallin}, we expect that the SW has positive dispersion with energy that increases monotonically with wave vector reaching a large wave vector asymptotic limit of $E^{*}_{\rm Z}$, the sum of $E_{\rm Z}$ and the Coulomb energy $E_{\rm C}$ required to create a particle-hole pair with reversed spin.\par
Here we report that CF ferromagnets exhibit a fundamentally different behavior. Resonant inelastic light scattering (ILS) measurements reveal the presence of excitations below the bare Zeeman energy. Quantitative interpretations based on CF theory reveal that the new modes are what we call spin-flip excitons (SFEs), namely excitations in which a CF quasiparticle transitions to a spin-reversed level with a lower $\Lambda$L quantum number. These are energetically more favorable than the ordinary ferromagnetic spin-wave excitations (which conserve the $\Lambda$L index) because of the associated lowering of the CF cyclotron energy. The SFEs produce one or more deep roton minima at finite wave vectors, and complex mode mixing creates a negative dispersion of the SW at small wave vectors. This behavior is fundamentally different from the dispersion of the conventional SW of electron ferromagnets.\par
The physics of spin conserving neutral excitations has been studied extensively in the past, resulting in the observation of dispersions of the neutral excitations of several FQHE states exhibiting roton minima \cite{Pinczuk93,Davies97,Cyrus05,Kukushkin09}. These observations are well explained in terms of spin conserving CF excitations across a single $\Lambda$L \cite{Dev92,Scarola00,Majumder09}. Excitations across multiple $\Lambda$L's have recently been observed \cite{Rhone11}. Relatively little attention had been paid so far to the spin-reversed excitations \cite{Mandal01,Dujovne03,Dujovne05}.
Many experimental studies have investigated the spin polarization of various FQHE ground state as well as spin phase transitions as a function of the Zeeman energy \cite{Eisenstein90,Du95,Kukushkin99,Verdene07}, analyzing their findings in terms of the CF $\Lambda$L transitions \cite{Park98}. While the present work deals with a different physics, namely, the spin-reversed excitations of the fully spin-polarized FQHE states, the appearance of roton minima below the bare Zeeman energy can be considered as a precursor of an instability of the ferromagnetic state at sufficiently low Zeeman energies.\par

\begin{figure}[h]
\centering
\includegraphics[width=0.5\textwidth]{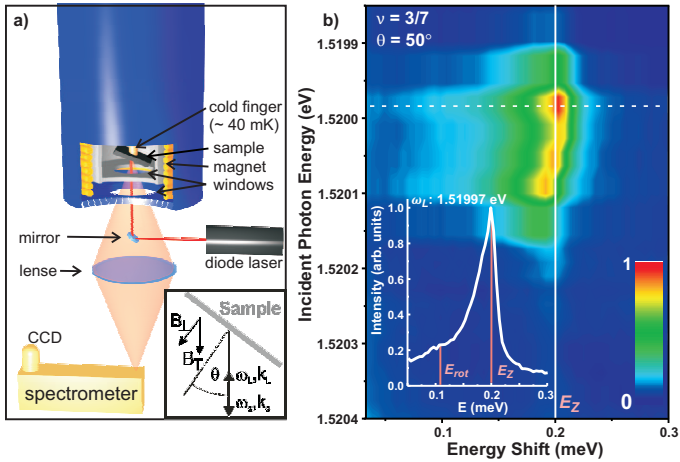}
\caption{Inelastic light scattering experiments. a) Experimental setup. b) Color plot of ILS intensities of spin-reversed excitations at $\nu$ = 3/7 as function of the exciting photon energy $E_{\rm L}$. The tilt angle is $\theta = 50^\circ$ and $B_{\rm tot}$ = 8 T. The inset shows spectra of the SW mode at $E_{\rm Z}$ and of sub-$E_{\rm Z}$ excitations for $E_{\rm L}$ = 1.51997 eV (indicated as a dashed line).}
\label{fig:fig1}
\end{figure}

The experiments were performed on an ultra high quality two-dimensional electron system confined in a single 33 nm wide asymmetric GaAs quantum well. Free electron density and mobility are $n = 5.35 \cdot 10^{10}$ cm$^{2}$ and $\mu = 7.2 \cdot 10^{6}$ cm$^{2}$/Vs, respectively, at $T$ = 300 mK after illumination. ILS measurements were performed in a $^{3}$He-$^{4}$He dilution refrigerator at a base temperature of 45 mK that is equipped with a 17 T superconducting magnet and windows for direct optical access [see Fig. \ref{fig:fig1} a)] The sample was placed with an angle $\theta$ between sample normal and external $B$-field as sketched in the inset of Fig. \ref{fig:fig1} a).
The resonant enhancement of the ILS was achieved by tuning the incident photon energies $E_{\rm L} = \hbar \omega_{\rm L}$ to be close to the fundamental optical gap of the GaAs QW \cite{Pinczuk93,Davies97,Jusserand00}. For ILS measurement of spin-reversed excitations the linearly polarized incoming light was cross polarized to the scattered light \cite{Dujovne03}. The spectra were acquired with a double monochromator and recorded with multichannel optical detection. The backscattering geometry employed in our experiments enables a momentum transfer $q_{0} = k_{\rm L} - k_{\rm S} = (2\omega_{\rm L}/c)sin\theta$, where $k_{\rm L(S)}$ is the in-plane component of the incident (scattered) photon. For our experiments, $q_{0}l \approx 0.09$ for $\theta = 30^\circ$ and $q_{0}l \approx 0.135$ for $\theta = 50^\circ$, where the magnetic length is $l = [eB_{\perp}/(\hbar c)]^{1/2}$.\par
Figure \ref{fig:fig1} b) displays a color plot of ILS intensities at $\nu = 3/7$ with $\theta = 50^\circ$ as a function of incident photon energies. The intensity of the long wavelength SW mode at $E_{\rm Z}$~=~0.2~meV is resonantly enhanced for $E_{\rm L}$ = 1.51997~meV.
\begin{figure}[b]
\centering
\includegraphics[width=0.4\textwidth]{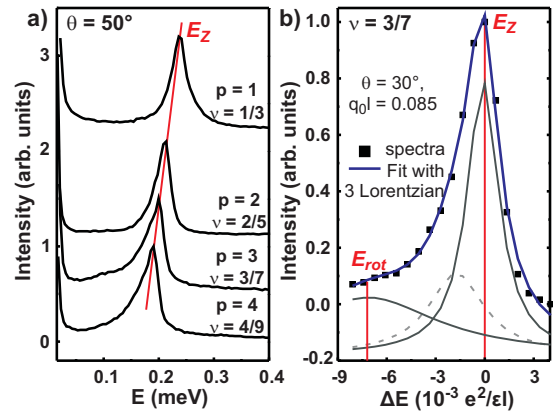}
\caption{Inelastic light scattering spectra in the energy range of lowest spin-reversed excitations. a) Results for $1/3 \leq \nu \leq 4/9$ ($1 \leq p \leq 4$) at $\theta =50^\circ$. The tail below $E_{\rm Z}$ increases with increasing $\Lambda$L number $p$. At $\nu =1/3$ only a high energy tail is observable. b) Spectra at $\nu = 3/7$ for $\theta =30^\circ$ (black squares). The blue line is a fit with three individual Lorentzians shown in gray.}
\label{fig:fig2}
\end{figure}
A low energy tail of spin-reversed excitations and the development of a new mode below $E_{\rm Z}$ (marked with a red line) are clearly visible. Similar modes below $E_{\rm Z}$ are observed at $\nu$ = 2/5 and 4/9 (Fig. \ref{fig:fig2} a)) at both $\theta = 30^\circ$ and $50^\circ$, although they are less pronounced for $30^\circ$ (Fig. \ref{fig:fig4}). The tail below $E_{\rm Z}$ becomes more enhanced with increasing CF filling factor $p$ as clearly visible in Fig. \ref{fig:fig2} a) for $\theta = 50^\circ$. No sub-Zeeman energy tail is seen at $\nu$ = 1/3. The low energy tail indicates the presence of spin-reversed modes below the Zeeman energy. This is further investigated by empirical fits to the resonant ILS spectra with two Lorentzians at $\nu$ = 2/5, three Lorentzians at $\nu$ = 3/7 and $\nu$ = 4/9 ($\theta = 30^\circ$) and four Lorentzians at $\nu$ = 4/9 ($\theta = 50^\circ$), which reproduce the experimental data very well, as demonstrated for $\nu$ = 3/7 ($\theta = 30^\circ$) in Fig. \ref{fig:fig2} b). In the empirical fits one Lorentzian is always centered at the bare Zeeman energy,  one or more Lorentzians is required to account for new excitations and one additional (dashed line) is needed to account for ILS intensities below $E_{\rm Z}$. The energies at the centers of the Lorenzians determined from the best fits are given in the last two columns in Table \ref{tab:tab1}. 
\begin{table}[b]
\centering
\begin{tabular}{cccccc}\hline\hline
$\nu$&mode&$ql$&$\Delta E_{\rm theory}$&$\Delta E_{30^\circ}$&$\Delta E_{50^\circ}$\\
 & & &$10^{-3}$($e^{2}/\epsilon l$)&$10^{-3}$($e^{2}/\epsilon l$)&$10^{-3}$($e^{2}/\epsilon l$)\\\hline\hline
2/5&$E_{\rm rot}$&0.373&-1.39&-0.98&-1.49\\
3/7&$E_{\rm rot}$&0.638&-7.48&-7.43&-6.20\\
4/9&$E^{1}_{\rm rot}$&0.63&-8.05&-6.20&-7.20\\
 &$E_{\rm max}$&1.134&-2.96&-2.80&-3.20\\
 &$E^{2}_{\rm rot}$&1.533&-5.69& &
\\\hline\hline
\end{tabular}
\caption{Momenta and energies of rotons and maxons modes from calculation and from Lorentzian fits to the experiment for both $\theta = 30^\circ$ and $50^\circ$. }
\label{tab:tab1}
\end{table}
While the empirical fits identify the major components of the ILS spectra of low-lying reversed-spin excitations, in the following we show that the observed modes are explained qualitatively and quantitatively as the spin-flip excitons of composite fermions. We note that the observation of a spin-wave mode indicates nonzero spin polarization. The spin phase diagram of FQHE determined previously both experimentally and theoretically indicates that our FQH system is in the region with full spin polarization \cite{Du95,Kukushkin99,Park98}. 
\begin{figure}[b]
\centering
\includegraphics[width=0.5\textwidth]{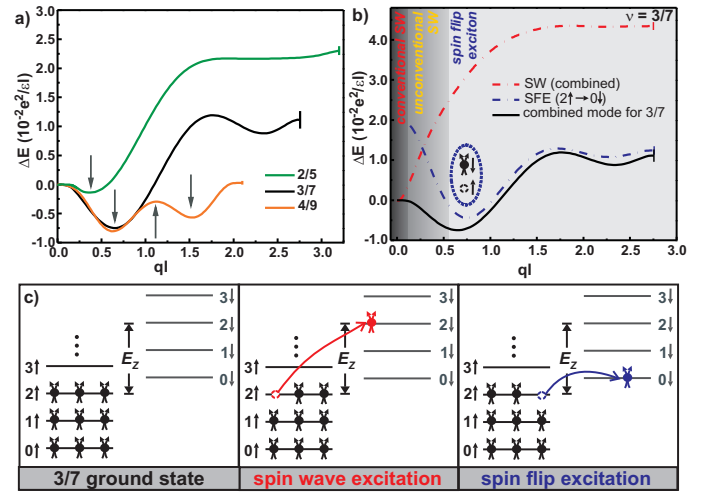}
\caption{Calculated wave vector dispersions. a)  The lowest spin-reversed excitation at $\nu$ =2/5, 3/7 and 4/9 for electron density $n = 5 \cdot 10^{10}$ cm$^{2}$ and QW width of 35 nm. Mode coupling is included as described in the text. The error bar at the end of each curve represents the typical statistical error in the energy determined from the Monte Carlo method. b) Wave vector dispersion at $\nu$ =3/7.  The dash-dotted red curve is for the lowest SW excitations without coupling to SFE modes. The blue dash-dotted line is the spin-flip exciton 2$\uparrow$ $\rightarrow$ 0$\downarrow$ without coupling to SW modes.  The black solid curve, same as in (a) for 3/7, incorporates mode couplings as described in the text. An illustration of the SFE is shown. c) $\Lambda$L scheme for fully spin-polarized QHE state at $p$ = 3 ($\nu$ = 3/7, 3/5). SW transition from 2$\uparrow$ $\rightarrow$ 2$\downarrow$ (center panel) and SFE transition from 2$\uparrow$ $\rightarrow$ 0$\downarrow$ (right panel) are outlined.}\label{fig:fig3}
\end{figure}
To illustrate the essential physics, we depict the spin-polarized CF ground state in Fig. \ref{fig:fig3} c) for $p$ = 3 ($\nu$ = 3/7, 3/5), for which the lowest three $\Lambda$Ls 0$\uparrow$, 1$\uparrow$ and 2$\uparrow$ are fully occupied. Various excitations are possible \cite{Mandal01,Park98}. The excitations that conserve the $\Lambda$L index (0$\uparrow$ $\rightarrow$ 0$\downarrow$,1$\uparrow$ $\rightarrow$ 1$\downarrow$,2$\uparrow$ $\rightarrow$ 2$\downarrow$) participate in the formation of the conventional SW excitation. The $\Lambda$L scheme reveals non-SW excitations, namely, the SFEs that change the $\Lambda$L index and simultaneously reverse spin. $\Lambda$L lowering SFEs are 2$\uparrow$ $\rightarrow$ 0$\downarrow$,2$\uparrow$ $\rightarrow$ 1$\downarrow$,1$\uparrow$ $\rightarrow$ 0$\downarrow$. Finally, the excitations that raise the $\Lambda$L index (such as 0$\uparrow$ $\rightarrow$ 1$\downarrow$,1$\uparrow$ $\rightarrow$ 2$\downarrow$,2$\uparrow$ $\rightarrow$ 3$\downarrow$) will not be considered below, because they are not relevant for the low energy physics of interest.
\par
\begin{figure}
\includegraphics[width=0.3\textwidth]{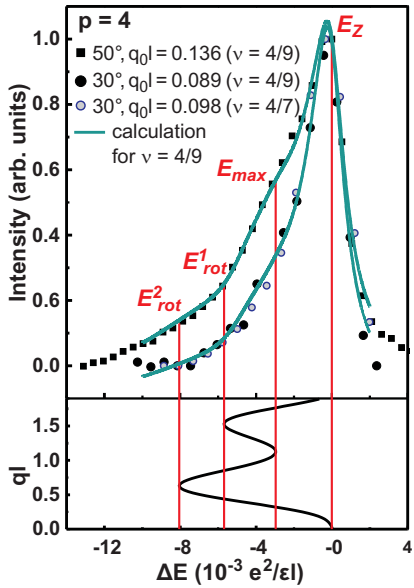}
\caption{Upper panel: Spectra at $\nu = 4/9$ for $\theta =50^\circ$ (solid squares) and $\theta =30^\circ$ (solid circles) and at $\nu = 4/7$ for $\theta =30^\circ$ (open blue circles). The solid lines are calculated ILS intensities for $\nu = 4/9$ at $\theta = 50^\circ$ and $30^\circ$. Lower panel: Calculated wave vector dispersion for $\nu = 4/9$.} \label{fig:fig4}
\end{figure}
This physical understanding suggests that the $\Lambda$L lowering SFEs, which are available for $p \geq 2$, may have energy below $E_{\rm Z}$, and it is tempting to associate them with the observed sub-Zeeman excitations. While this physics naturally explains the absence of negative energy excitations at 1/3, its unambiguous confirmation requires a quantitative comparison with theory, which in turn necessitates a consideration of the mixing between these excitations due to the residual inter-CF interaction. Such calculations are possible within the well developed formalism of the CF theory, which has been found to be quantitatively successful in describing the spin phase diagram of the FQHE states. We calculate the dispersions of the true eigenmodes for $\nu$ = 2/5, 3/7, and 4/9 by the method of CF diagonalization, in which the Coulomb interaction is diagonalized in the space of excitations listed above. We include the effective renormalization of the interaction due to finite quantum well width as explained in Refs. \onlinecite{Majumder09,Rhone11}. Landau-level mixing and residual disorder has been neglected. The dispersions are computed with up to 136 particles by methods described in the literature \cite{Mandal01} and represent the thermodynamic limit. The predicted dispersion for the lowest spin-reversed modes for 2/5, 3/7, and 4/9 are plotted in Fig. \ref{fig:fig3} a). The typical statistical error in the energy resulting from the Monte Carlo method is given at the end of each dispersion curve.\par
A striking and unusual feature is that the SW dispersion has a negative curvature at small wave vectors. A second notable aspect is that at 2/5, 3/7, and 4/9 one or more minima and maxima, called rotons and maxons, appear below $E_{\rm Z}$. The lowest spin-reversed branch is a mixture of the SW and SFE excitations, and changes its dominant character as a function of the wave vector. As shown in Fig. \ref{fig:fig3} b) exemplarily for $\nu$ = 3/7, it is predominantly 2$\uparrow$ $\rightarrow$ 0$\downarrow$ SFE-like for $ql > 0.8$, and converges into the SW in the $ql \rightarrow 0$ limit. At intermediate wave vectors, however, it is a more complicated admixture of the SW and several SFE modes. Remarkably, the energies of all individual modes are positive in this intermediate wave vector range, whereas the energy of the combined mode is negative. Nonetheless, the 2$\uparrow$ $\rightarrow$ 0$\downarrow$ SFE mode captures all the rotons and maxons, with other SFEs resulting in further lowering of the energy but without producing new qualitative structures.\par
The change in the photon wave vector is in general small compared to $1/l$. However, as it is well appreciated, disorder allows (discussed below in more detail) coupling of ILS to the excitations at wave vectors comparable to $1/l$. The coupling is strongest to the SFEs near the critical points in the dispersion, i.e. the rotons and the maxons, because of a singularity in the density of states at those energies \cite{Cyrus05}. The theoretically predicted energies of the SFE rotons and maxons, obtained without any adjustable parameters, are shown in Table \ref{tab:tab1}. The excellent agreement between the calculated and the measured energies gives a quantitative confirmation of our physical assignment of these modes. For a more detailed comparison, we model the ILS intensity $I(E,q_{0})$ by the phenomenological expression \cite{Marmorkos92}
\begin{equation}
     I(E,q_0) \propto \int f(q,q_0) \rho (E,q) dq,
\end{equation}
where $q_{0}$ is the photon wave vector transfer, and $f(q, q_{0}) = (1/2)q_{b}/[(q - q_{0})^{2} + q_{b}^{2}]$ is a Lorenzian that incorporates the physics of the breakdown of wave vector conservation with $q_{b}$ setting its scale. The quantity $\rho(E,q) = (\Gamma/2)/[(E - E(q))^{2} +\Gamma^{2}]$ is the mode response function with $\Gamma$ being a measure of broadening of the theoretical dispersion $E(q)$ due to experimental resolution as well as disorder.
In this analysis both $\Gamma$ and $q_{b}$ are treated as adjustable parameters. The best fits are obtained for $\Gamma = 8 \cdot 10^{4}$ $E_{\rm C}$ and $q_{b} \approx 1.45$ $l$. The results (solid lines) of our phenomenological model are compared with the experimentally observed spectra (scatter) in Fig. \ref{fig:fig4} for $\nu$ = 4/9 ($\theta = 30^\circ, 50^\circ$). The excellent agreement provides further verification of the physics presented above. Figure \ref{fig:fig3} c) shows that the observed ILS spectrum at 4/7 closely resembles that at 4/9, which is nicely consistent with the fact that both states have $p = 4$ filled $\Lambda$L's of composite fermions (albeit with opposite $B_{\rm eff}$).  Similar agreements are found between the ILS spectra at 2/3 and 2/5, and between those at 3/5 and 3/7. The physics of the spin-reversed excitations is thus governed by the CF $\Lambda$L number $p$ and not the electronic filling factor $\nu$. A direct comparison between theory and experiment for the $p/(2p - 1)$ states is not possible due to technical reasons that render an evaluation of $E(q)$ computationally prohibitively expensive.\par

\textbf{Acknowledgments} - The work at Columbia was supported by the National Science Foundation (NSF) (DMR-03-52738, DMR-0803445, and CHE-0641523), by the U.S. Department of Energy (DOE) (DE-AI02-04ER46133). The work at Princeton was partially funded by the Gordon and Betty Moore Foundation as well as the NFS MRSEC Program through the Princeton Center for Complex Materials (DMR-0819860). U.W. was partially supported by the Alexander von Humboldt Foundation. J.K.J. was supported in part by the NSF (DMR-1005536). The computation was performed using the cluster of the Department of Theoretical Physics, Indian Association for the Cultivation of Science.


\begin{thebibliography}{23}
\expandafter\ifx\csname natexlab\endcsname\relax\def\natexlab#1{#1}\fi
\expandafter\ifx\csname bibnamefont\endcsname\relax
  \def\bibnamefont#1{#1}\fi
\expandafter\ifx\csname bibfnamefont\endcsname\relax
  \def\bibfnamefont#1{#1}\fi
\expandafter\ifx\csname citenamefont\endcsname\relax
  \def\citenamefont#1{#1}\fi
\expandafter\ifx\csname url\endcsname\relax
  \def\url#1{\texttt{#1}}\fi
\expandafter\ifx\csname urlprefix\endcsname\relax\def\urlprefix{URL }\fi
\providecommand{\bibinfo}[2]{#2}
\providecommand{\eprint}[2][]{\url{#2}}


\bibitem{Tsui82}
D.~C.~Tsui, H.~L.~Stormer, and A.~C.~Gossard, Phys. Rev. Lett., {\bf 48}, 1559 (1982).

\bibitem{Jain89}
J.~K.~Jain, Phys. Rev. Lett., {\bf 63}, 199 (1989).

\bibitem{Kallin}
C.~Kallin and B.~I.~Halperin, Phys. Rev. B {\bf 30}, 5655 (1984).

\bibitem{Pinczuk93}
A.~Pinczuk, B.~S.~Dennis, L.~N.~Pfeiffer, and K.~West, Phys. Rev. Lett. {\bf 70}, 3983 (1993).

\bibitem{Davies97}
H.~D.~M.~Davies, J.~C.~Harris, J.~F.~Ryan, and A.~J.~Turberfield, Phys. Rev. Lett. {\bf 78}, 4095 (1997).

\bibitem{Cyrus05}
C.~F.~Hirjibehedin, I.~Dujovne, A.~Pinczuk, B.~S.~Dennis,
L.~N.~Pfeiffer, and K.~W.~West, Phys. Rev. Lett. {\bf 95}, 066803 (2005).

\bibitem{Kukushkin09}
I.~V.~Kukushkin,  J.~H.~Smet, V.~W.~Scarola, V.~Umansky,  and
K.~v.~Klitzing, Science {\bf 324}, 1044 (2009).

\bibitem{Dev92}
G.~Dev and J.~K.~Jain, Phys. Rev. Lett. {\bf 69}, 2843 (1992).

\bibitem{Scarola00}
V.~W.~Scarola, K.~Park, and J.~K.~Jain, Phys. Rev. B {\bf 61}, 13064 (2000).

\bibitem{Majumder09}
D.~Majumder, S.~S.~Mandal, and J.~K.~Jain, Nature Physics {\bf 5}, 403 (2009).

\bibitem{Rhone11}
T.~D.~Rhone, D.~Majumder, B.~S.~Dennis, C.~Hirjibehedin, I.~Dujovne,
J.~G.~Groshaus, Y.~Gallais, J.~K.~Jain, S.~S.~Mandal, A.~Pinczuk,
et~al., Phys. Rev. Lett. {\bf 106}, 096803 (2011).

\bibitem{Mandal01}
S.~S.~Mandal and J.~K.~Jain, Phys. Rev. B {\bf 63}, 201310(R) (2001).

\bibitem{Dujovne05}
Irene Dujovne, A. Pinczuk, Moonsoo Kang, B. S. Dennis, L. N.
Pfeiffer, and K. W. West Phys. Rev. Lett. {\bf 95}, 056808 (2005).

\bibitem{Dujovne03}
I.~Dujovne, A.~Pinczuk, M.~Kang, B.~S.~Dennis, L.~N.~Pfeiffer, and
K.~W.~West, Phys. Rev. Lett. {\bf 90},  036803 (2003).

\bibitem{Du95} R.R. Du, A. S. Yeh, H. L. Stormer, D. C. Tsui, L. N. Pfeiffer, and K. W. West, Phys. Rev. Lett. {\bf 75}, 3926 (1995).

\bibitem{Eisenstein90} J. P. Eisenstein, H. L. Stormer, L. N. Pfeiffer, and K. W West, Phys. Rev. B {\bf 41}, 7910 (1990).

\bibitem{Kukushkin99} I.V. Kukushkin, K.v. Klitzing and K. Eberl, Phys. Rev. Lett. {\bf 82}, 3665 (1999).

\bibitem{Verdene07} B. Verdene, J. Martin, G. Gamez, J. Smet, K. v. Klitzing, D. Mahalu, D. Schuh, G. Abstreiter, and A. Yacoby, Nature Physics {\bf 3}, 392 (2007).

\bibitem{Park98} K. Park and J. K. Jain, Phys. Rev. Lett. {\bf 80}, 4237 (1998).

\bibitem{Jusserand00}
B.~Jusserand, M.~N.~Vijayaraghavan, F.~Laruelle, A.~Cavanna, and  B.~Etienne, Phys. Rev. Lett. {\bf 85}, 5400 (2000).


\bibitem{Marmorkos92} I. K. Marmorkos and S. Das Sarma, Phys. Rev. B {\bf 45}, 13396 (1992).

\end{thebibliography}

\pagebreak

\end{document}